\documentclass[submission,copyright,creativecommons]{eptcs}
\providecommand{\event}{ACL2 Workshop 2022}
\usepackage{underscore, cjhebrew, latexsym,pifont,amsmath,amstext,amssymb,amsfonts,makeidx,url}
\title{Properties of the Hebrew Calendar}
\author{David M.~Russinoff
\email{david@russinoff.com}
}
\def\titlerunning{Properties of the Hebrew Calendar}
\def\authorrunning{D.M. Russinoff}

\begin{document}
\maketitle

\begin{abstract}
We describe an ACL2 program that implements the Hebrew calendar and the formal verification of several of its properties, including the
critical result that the algorithm that determines the placement of the new year ensures that the length of every year belongs to a small
set of admissible values.  These properties have been accepted for many centuries without the benefit of explicit proof, in spite of
subtleties in the underlying arguments. For the sake of accessibility to a broad audience, the program is coded in Restricted Algorithmic
C (RAC), a simple language consisting of the most basic constructs of C, for which an automatic translator to the ACL2 logic has been
implemented.  While RAC is primarily intended for modeling arithmetic hardware designs, this novel application provides a relatively
simple illustration of the language and the translator.
\end{abstract}

\section{Introduction}

Lunisolar calendars are complicated by the definitional requirements of tracking both the moon's orbit about the earth and that
of the earth about the sun.  The first of these objectives is achieved by the same scheme that underlies a pure lunar calendar: the
year is partitioned into months that are designed to coincide as closely as possible with the cycle of phases of the moon, which is
delimited by successive {\it lunar conjunctions}.  A conjunction occurs when the moon lies on the plane perpendicular to that of the
earth's orbit that passes through the earth and the sun, and therefore, the earth lies on the dark side of the moon.  The duration of
this cycle is subject to significant variation (owing mainly to variation in the earth's angular velocity with respect to the sun), but
its mean value, known as the {\it synodic month}, has been estimated to be 29.53059 days.  Since the mean period of revolution of the
earth, known as the {\it tropical year}, is 365.2422 days, the second objective requires that the average number of months in a year be
close to the ratio $365.2422/29.53059 \approx 12.36827$.  This is effected by interleaving 13-month {\it leap} years with 12-month
{\it common} years at an appropriate frequency.

The lunisolar nature of the Hebrew calendar is dictated by (a) the Biblical ritual of sanctifying the new moon at the beginning of
each month and (b) traditional agriculture components of various Jewish holidays, which require alignment with the seasons.  In
ancient times, both requirements were implemented dynamically through empirical observation: the beginning of a month was declared upon
sighting of the new moon, and an extra month was inserted into the year as required to ensure that Passover, the ``Festival of spring'',
would occur during the barley season.  Such information was traditionally spread through the diaspora by messenger.  This process
ultimately proved impractical in the face of Roman persecution and was replaced by a fixed calendar based on astronomical calculations.

The fixed Hebrew calendar was instituted in 359 C.E. (Hebrew year 4119) and continued to evolve through the $9^{\rm th}$ century.  It
remains in widespread use today by the world's Jewish population, mainly for the purpose of tracking holidays and {\it yahrzeits} (death
anniversaries).  The calendar's design, which remains poorly understood even by those who regularly rely on it, is the subject of this note.
A separate exposition, focusing on the history of the calendar and some of the underlying astronomy, is available at
{\tt www.russinoff.com/cal\-endar/calendar.pdf}.  Here we shall describe an ACL2 program that defines the calendar and enables the formal
verification of several of its critical properties, including the critical result that the algorithm that determines the placement of the
new year ensures that the length of every year belongs to a small set of admissible values.  These properties have been accepted for many
centuries without the benefit of explicit proof, in spite of subtleties in the underlying arguments.

The program is hand-coded in {\it Restricted Algorithmic C (RAC)}\cite{workshop, el}, a simple language intended primarily for the modeling
and formal verification of arithmetic hardware designs.  RAC consists of the most basic constructs of C augmented by the C++ register class
templates of Algorithmic C\cite{ac}, which provide the bit manipulation primitives of a hardware description language.  Various restrictions
on the language are designed to promote a functional programming paradigm, thereby facilitating its automatic translation to
the logic of ACL2.  While this calendar program does not utilize the specialized features of RAC, our objective is an executable model that
is easily understood and accessible to a broad audience beyond the ACL2 community.  This novel application also provides a relatively simple
illustration of the language and the translator.  The proof script resides in the ACL2 directory {\tt books/workshops/2022/russinoff-calendar/}.
The RAC program and its ACL2 translation reside in the files {\tt calendar.cpp} and {\tt calendar.lisp}; the proof script is in {\tt proof.lisp}.

\section{Approximating Lunar Conjunctions}\label{moladot}

For the purpose of calendrical calculations, a Hebrew day begins at 6 PM in Jerusalem and the time of day is measured in hours and parts,
where an hour comprises 1080 parts.  For example, 4 hours, 0 parts, which we denote as 4:0, is 10 PM and 18:270 is 15 minutes after noon.
The days of the week are numbered, beginning at 6 PM on Saturday.  We shall adhere to the usual abuse of notation in using English names to
refer to Hebrew days, e.g., ``Monday'' is understood to refer to the second day of the week, which is actually the 24-hour period that
begins at 6 PM on Sunday.

The modern Hebrew calendar inherits a number of elements from those of the Babylonians and Greeks.  One is a remarkably accurate
estimate of the synodic month as 29 days, 12 hours, and 793 parts, approximately 29.53059 days, which we shall denote
as ${\cal L}$.  Another is the {\it Metonic cycle}, which is based on the observation that the ratio of the tropical year to the synodic
month is closely approximated by the ratio $\frac{235}{19} \approx 12.36842$.  This gives rise to a 19-year cycle consisting of 12
common years and 7 leap years, with a total of $12\cdot 12 + 7\cdot 13 = 235$ months and an average of $235/19$ months/year.  Thus, 
by convention, a Hebrew year $y$ is a leap year if $y \bmod 19 \in \{0, 3, 6, 8, 11, 14, 17\}$.
These definitions are formalized by the RAC predicates
\begin{small}
\begin{verbatim}
  bool leap(uint year) {
    uint m = year % 19;
    return m == 0 || m == 3 || m == 6 || m == 8 ||
           m == 11 || m == 14 || m == 17;
  }
\end{verbatim}
\end{small}
and
\begin{small}
\begin{verbatim}
  bool common(uint year) {return !leap(year);}
\end{verbatim}
\end{small}
and the function
\begin{small}
\begin{verbatim}
  uint monthsInYear(uint year) {return leap(year) ? 13 : 12;}
\end{verbatim}
\end{small}
each of which takes a positive integer argument.

The 12 months of a common year are Tishri, Cheshvan, Kislev, Tevat, Shevat, Adar, Nisan, Iyar, Sivan, Tammuz, Av, and Elul.  In view
of the length of the synodic month, the calendar months are tentatively set to alternate between 30 and 29 days, beginning with
the 30-day month of Tishri.  Since the resulting average length is a slight underestimate, the extra month of a leap year is 
assigned 30 days.  This is designated as an additional month of Adar, named Adar I and inserted before the 29-day Adar, which becomes
Adar II.  This scheme (which will be refined in Section~\ref{rh}), would result in common and leap years of lengths
$6 \times 29 + 6 \times 30 = 354$ and and $354 + 30 = 384$, respectively.

The next step in the design of the calendar is an estimate of the time of the lunar conjunction that marks the beginning of each month.
This estimate is called a {\it molad} (from the root of the Hebrew word for ``birth''), with plural {\it moladot}.  The molad of a year
will be understood to be the molad of Tishri of that year.  For this purpose, the molad of the year 2 is taken to be precisely 14 hours (8 AM)
on a Friday.  According to tradition, this occurred on the sixth day of creation, which is assumed to have taken place during the final week
of the year 1.  As the story goes, apparently based on the long-held erroneous belief that the crescent moon becomes visible 6 hours after
the conjunction, the new-born Adam witnessed the first new moon at 2 PM on that day.  The actual origin of {\it Molad Adam} is unknown,
having been obscured by the mythology, but evidence suggests~\cite{stern} that it was computed by rounding a value derived from tables
recorded by Ptolemy in the $2^{\rm nd}$ century C.E.

As a convenient point of reference, the (imaginary) molad of the year 1 is computed by subtracting $12\cdot {\cal L}$ from Molad
Adam, resulting in 5:204 on the second day of the week.  This is known as {\it Molad Beharad}, a transliteration
of the Hebrew \<bhrd>, which is derived from the conventional assignment of numerical values to letters (\<b> = 2, \<h> = 5, \<r> = 200,
\<d> = 4).  We find that our analysis is facilitated by introducing an {\it absolute calendar}, which may be considered as having
a single year with a single month of infinite length.  Since Beharad falls on the second day of the week, it is natural to set the first
day of this calendar to the preceding day, so that the day of the week of an absolute date is computed by
\begin{small}
\begin{verbatim}
  uint dayOfWeek(uint day) {return day % 7;}
\end{verbatim}
\end{small}

The program defines the type
\begin{small}
\begin{verbatim}
  struct Moment {uint day; uint hour; uint part;};
\end{verbatim}
\end{small}
to encode a duration of $d$ days, $h$ hours, and $p$ parts.  Following C syntax, we shall represent such a duration as $\{d, h, p\}$, e.g.,
${\cal L} = \{29, 12, 793\}$.\footnote{This convention relies on context to distinguish such an object from a set of three elements.}  This
is encoded as
\begin{small}
\begin{verbatim}
  const Moment Lunation = {29, 12, 793};
\end{verbatim}
\end{small}
The same {\tt struct} type and informal notation may represent a moment in time, using the absolute date for $d$.  Thus, Molad
Beharad is ${\cal B} = \{2, 5, 204\}$, or
\begin{small}
\begin{verbatim}
  const Moment Beharad = {2, 5, 204};
\end{verbatim}
\end{small}
The RAC translator converts these structures to alists and makes use of the access and setting functions {\tt AG} and {\tt AS}, e.g.,
\begin{small}
\begin{verbatim}
  (DEFUND LUNATION NIL (AS 'PART 793 (AS 'HOUR 12 (AS 'DAY 29 NIL))))
\end{verbatim}
\end{small}
the value of which is {\tt ((DAY . 29) (HOUR . 12) (PART . 793))}.

The addition of these objects and multiplication by a positive integer are defined in the natural way.  Thus, Molad Adam is 
\[
{\cal B} + 12\cdot {\cal L} = \{356, 14, 0\}.
\]
These computations are formalized by the RAC functions
\begin{small}
\begin{verbatim}
  Moment addTime(Moment x, Moment y) {
    Moment sum;
    uint sumParts = x.part + y.part;
    sum.part = sumParts % 1080;
    uint sumHours = x.hour + y.hour + sumParts/1080;
    sum.hour = sumHours % 24;
    sum.day = x.day + y.day + sumHours/24;
    return sum;
  }
\end{verbatim}
\end{small}
and
\begin{small}
\begin{verbatim}
  Moment mulTime(uint m, Moment x) {
    Moment prod;
    uint prodParts = m * x.part;
    prod.part = prodParts % 1080;
    uint prodHours = m * x.hour + prodParts/1080;
    prod.hour = prodHours % 24;
    prod.day = m * x.day + prodHours/24;
    return prod;
  }
\end{verbatim}
\end{small}
Thus,
\begin{small}
\begin{verbatim}
  Moment molad(uint year) {
    // Compute the total number of months in all preceding years, 
    // multiply by the lunation period, and add the product to Beharad:
    uint priorMonths = 0;
    for (uint y=1; y<year; y++) {
      priorMonths += monthsInYear(y);
    }
    return addTime(Beharad, mulTime(priorMonths, Lunation));
  }
\end{verbatim}
\end{small}
For example, the molad of the year of the $17^{\rm th}$ ACL2 Workshop is computed as
\begin{eqnarray*}
{\tt molad(5782)} & = & {\tt addTime(Beharad, mulTime(71489, Lunation))}\\
& = & {\tt \{2111469, 5, 497\}}.
\end{eqnarray*}
Note that a RAC {\tt for} loop generates an auxiliary recursive ACL2 function, which in the case of the function {\tt molad} computes
the total number of months that occur between two given years:
\begin{small}
\begin{verbatim}
  (DEFUND MOLAD-LOOP-0 (Y YEAR PRIORMONTHS)
          (DECLARE (XARGS :MEASURE (NFIX (- YEAR Y))))
          (IF (AND (INTEGERP Y)
                   (INTEGERP YEAR)
                   (< Y YEAR))
              (LET ((PRIORMONTHS (+ PRIORMONTHS (MONTHSINYEAR Y))))
                   (MOLAD-LOOP-0 (+ Y 1) YEAR PRIORMONTHS))
              PRIORMONTHS))

  (DEFUND MOLAD (YEAR)
          (LET* ((PRIORMONTHS 0)
                 (PRIORMONTHS (MOLAD-LOOP-0 1 YEAR PRIORMONTHS)))
                (ADDTIME (BEHARAD)
                         (MULTIME PRIORMONTHS (LUNATION)))))
\end{verbatim}
\end{small}
Establishing the properties of such functions naturally involves induction.  In this case, the following decomposition theorem is derived
directly from the definition, with no hints:
\begin{small}
\begin{verbatim}
  (defthmd molad-loop-decomp
    (implies (and (natp prior) (posp y) (posp k) (posp year) (<= y k) (<= k year))
             (equal (molad-loop-0 y year prior)
                    (molad-loop-0 k year (molad-loop-0 y k prior)))))
\end{verbatim}
\end{small}
Instantiation of the above result with $y \leftarrow 1$, $\mathit{year} \leftarrow y+1$, $k \leftarrow y$, and $\mathit{prior} \leftarrow 0$
yields the following recurrence formula:
\begin{small}
\begin{verbatim}
  (defthmd molad-next
    (implies (posp year)
             (equal (molad (1+ year))
                    (addtime (molad year)
                              (multime (monthsinyear year) (lunation))))))
\end{verbatim}
\end{small}
This formula significantly improves efficiency in computing a long sequence of moladot, as discussed in Section~\ref{program}.

\section{Determining the New Year}\label{rh}

{\it Rosh Hashanah}, the Jewish new year, is traditionally celebrated during the first two days of Tishri, but we shall use the term to refer
to Tishri 1.  A naive objective is to arrange for the molad of each year to occur on Rosh Hashanah.  This is not generally possible,
however, with month and year lengths as tentatively specified in Section~\ref{moladot}.  For example, the molad of a common year is separated
from that of the following year by an interval of $12\cdot {\cal L} = \{354, 8, 817\}$.  Consequently, depending on the time of day of the
first of these moladot, the absolute dates on which they occur may differ by either 354 or 355.  Similarly, since
$13\cdot {\cal L} = \{383, 21, 589\}$, the absolute dates of the moladot that begin and end a leap year may differ by either 383 or 384.  All
four of these year lengths may be accommodated by allowing the lengths of two specific months, Cheshvan and Kislev, to vary between 29 and 30,
as displayed in Figure~\ref{months}.  Thus, if the length of a common year is computed (from the absolute dates of the bounding moladot) to be
355, then the lengths of Cheshvan and Kislev are both set to 30.  Similarly, if the length of a leap year is found to be 383, then both
months are assigned 29 days.  The first case is called a {\it complete} common year, the second is a {\it defective} leap year, and a
year of length 354 or 384 is said to be {\it regular}.

\begin{figure}[h]

\setlength{\unitlength}{2mm}
\begin{picture}(35,45)(-7,0)
\put(0,0){\line(1,0){45}}
\put(0,3){\line(1,0){45}}
\put(0,6){\line(1,0){45}}
\put(0,9){\line(1,0){45}}
\put(0,12){\line(1,0){45}}
\put(0,15){\line(1,0){45}}
\put(0,18){\line(1,0){45}}
\put(0,21){\line(1,0){45}}
\put(0,24){\line(1,0){45}}
\put(0,27){\line(1,0){45}}
\put(0,30){\line(1,0){45}}
\put(0,33){\line(1,0){45}}
\put(0,36){\line(1,0){45}}
\put(0,39){\line(1,0){45}}
\put(0,39.5){\line(1,0){45}}
\put(0,42.5){\line(1,0){45}}
\put(3,1){\large{Elul}}
\put(3,4){\large{Av}}
\put(3,7){\large{Tammuz}}
\put(3,10){\large{Sivan}}
\put(3,13){\large{Iyar}}
\put(3,16){\large{Nisan}}
\put(3,19){\large{Adar (Adar II on leap year)}}
\put(3,22){\large{Adar I (leap year only)}}
\put(3,25){\large{Shevat}}
\put(3,28){\large{Tevat}}
\put(3,31){\large{Kislev}}
\put(3,34){\large{Cheshvan}}
\put(3,37){\large{Tishri}}
\put(0,39.5){\makebox(31,3){\large{Month}}}
\put(34,1){\large{29}}
\put(34,4){\large{30}}
\put(34,7){\large{29}}
\put(34,10){\large{30}}
\put(34,13){\large{29}}
\put(34,16){\large{30}}
\put(34,19){\large{29}}
\put(34,22){\large{30}}
\put(34,25){\large{30}}
\put(34,28){\large{29}}
\put(34,31){\large{30 or 29}}
\put(34,34){\large{29 or 30}}
\put(34,37){\large{30}}
\put(31,39.5){\makebox(14,3){\large{Days}}}
\put(0,0){\line(0,1){42.5}}
\put(31,0){\line(0,1){42.5}}
\put(45,0){\line(0,1){42.5}}
\end{picture}
\caption{Months of the year}\label{months}
\end{figure}

However, as the calendar evolved between the $4^{\rm th}$ and $9^{\rm th}$ centuries, a variety of constraints were imposed,
resulting in the possible delay of Rosh Hashanah by up to two days beyond the date of the molad.  This led to additional variation in
the length of a year.  We shall demonstrate, however, that these postponements, or {\it dechiyot}, effectively ensure that the
resulting length of a common year is always 353, 354, or 355, and that of a leap year is always 383, 384, or 385.  Consequently, the
month lengths listed in Figure~\ref{months} remain in force, but now with the understanding that either a common year or a leap year
may be defective (353 or 383), regular (354 or 384), or complete (355 or 385).  The four dechiyot are traditionally listed in the
following order:\medskip

\noindent
{\bf First Dechiyah}: {\it If the molad of Tishri occurs at or after noon, then Rosh Hashanah is postponed to the next day.}\medskip

The rationale for this postponement is far from clear.  A common explanation is that the rule is intended to ensure that the new moon
is visible in the sky at the beginning of Rosh Hashanah.  This is consistent with the belief mentioned earlier in connection with Molad
Adam, but is unrealistic since the moon rarely appears within 24 hours on either side of the molad.  An alternative reason is suggested
by Landau~\cite{landau}: this delay guarantees that the molad of every month always occurs before the end of the first day of the
month.  (Demonstration of this property requires some computation; see Section~\ref{program}.)

Whatever the motivation for this dechiyah, its simplest implementation is to replace the molad with the moment that occurs 6 hours
later, which we shall call the {\it delayed molad}, represented by the RAC function
\begin{small}
\begin{verbatim}
  Moment dmolad(uint year) {
    Moment sixHours = {0, 6, 0};
    return addTime(molad(year), sixHours);
  }
\end{verbatim}
\end{small}
This modification effectively bypasses the $1^{\rm st}$ dechiyah.  That is, we begin with the assumption that Tishri 1 is tentatively
scheduled for the day of the delayed molad and replace the usual statements of the remaining dechiyot with reformulations expressed
in terms of the delayed molad.\footnote{The decision to follow this path, which I considered when I first encountered the list of
dechiyot, became easier when I learned from Bromberg\cite{bromberg} that the same revision was proposed three centuries ago
by K.~F.~Gauss.}\medskip

\noindent
{\bf Second Dechiyah}: {\it If the delayed molad occurs on a Wednesday, Friday, or Sunday, then Rosh Hashanah is postponed to the next
day.}\medskip

The reasons most commonly cited for the prohibition of these three days involve other holidays later in the month of Tishri.  If Rosh
Hashanah were to fall on a Wednesday or Friday, then Yom Kippur, Tishri 10, would occur on a Friday or Sunday, and hence would be adjacent
to Shabbat (Saturday), with the untenable result of successive days of restricted activities.  If Rosh Hashanah were to fall on a Sunday,
then Hoshanah Rabbah, the $21^{\rm st}$ day of Tishri, would occur on Shabbat.  This must be avoided because the ritual observance of this
holiday (including the beating of willow branches against the floor, symbolizing the elimination of sin) constitutes work of the sort that
is prohibited on Shabbat.

If the $2^{\rm nd}$ dechiyah were applied in isolation, then some years would have inadmissible lengths. The remaining two dechiyot
are designed to circumvent this result.  The $3^{\rm rd}$ addresses the case of a common year and the $4^{\rm th}$ the case of a
leap year.  (The proof of correctness is deferred to Section~\ref{program}.)\medskip

\noindent
{\bf Third Dechihah}: {\it If the delayed molad of a common year occurs on a Tuesday at or later than 15:204, then Rosh Hashanah
is delayed to the following Thursday.}\medskip

\noindent
{\bf Fourth Dechihah}: {\it If the delayed molad following a leap year occurs on a Monday at or later than 21:589, then Rosh
Hashanah is delayed to the next day.}\medskip

A trivial consequence of this procedure is that Rosh Hashanah cannot occur on a Wednesday, Friday, or Sunday.  Note also that at most
one of the $2^{\rm nd}$, $3^{\rm rd}$, and $4^{\rm th}$ dechiyot can apply to any delayed molad.  The $3^{\rm rd}$ advances Rosh
Hashanah by two days, and each of the others by one.  Furthermore, if the date of the molad is different from that of the delayed
molad, then the latter must occur before midnight and neither the $3^{\rm rd}$ nor the $4^{\rm th}$ applies.  Thus, Rosh Hashanah is
delayed at most two days from the day of the true molad.

These computations are formalized as follows:

\begin{small}
\begin{verbatim}
  uint roshHashanah(uint year) {
    Moment dm = dmolad(year);
    uint day = dm.day;
    uint dw = dayOfWeek(day);
    if (dw == 1 || dw == 4 || dw == 6) {
      // 2nd dechiyah
      day++;
    }
    else if (dw == 3 && !earlier(dm, 15, 204) && !leap(year)) {
      // 3rd dechiyah
      day = day + 2;
    }
    else if (dw == 2 && !earlier(dm, 21, 589) && leap(year-1)) {
      // 4th dechiyah
      day++;
    }
    return day;
  }
  uint yearLength(uint year) {
    return roshHashanah(year+1) - roshHashanah(year);
  }
\end{verbatim}
\end{small}

\section{Conversion Between Calendars}

In order to establish confidence in the program, we define procedures that convert between Hebrew and Gregorian dates, which
may be checked against established sources.  A date of either sort is represented as an object of type

\begin{small}
\begin{verbatim}
  struct Date {uint day; uint month; int year;};
\end{verbatim}
\end{small}
Here the months of a Gregorian year or a common Hebrew year are encoded in order as the integers $1,\ldots,12$, and the extra month
of a Hebrew leap year, Adar I, is encoded as 13, although it occurs between months 5 (Shevat) and 6 (Adar II).  Under this encoding,
the molad of any month is readily derived from the molad of the year:
\begin{small}
\begin{verbatim}
  Moment monthlyMolad(uint month, uint year) {
    uint priorMonths;
    if (leap(year) && month >= 6) {
      priorMonths = month == 13 ? 5 : month;
    }
    else {
      priorMonths = month - 1;
    }
    return addTime(molad(year), mulTime(priorMonths, Lunation));
  }
\end{verbatim}
\end{small}
The length of a given month is determined by the length of the year in which it occurs:

\begin{small}
\begin{verbatim}
  uint monthLength(uint month, uint yearLen) {
    uint monLen;
    switch(month) {
    case 2: monLen = yearLen == 355 || yearLen == 385 ? 30 : 29; break;
    case 3: monLen = yearLen == 353 || yearLen == 383 ? 29 : 30; break;
    default: monLen = month % 2 == 0 ? 29 : 30;
    }
    return monLen;
  }
\end{verbatim}
\end{small}

The absolute date corresponding to a given Hebrew date is computed by the function {\tt h2a} as the sum of three integers:

\begin{itemize}
\item [(1)] The absolute date of the last day of the preceding year, computed as {\tt roshHashanah(year - 1)};
\item [(2)] The number of days of the year that precede the month in which the date occurs, computed by summing the values of the function
{\tt monthLength} applied to the preceding months (which requires separate computations for common and leap years);
\item [(3)] The day of the month.
\end{itemize}
Thus, we have the following definition:

\begin{small}
\begin{verbatim}
  uint h2a(Date date) {
    // Compute number of days of date.year that precede date.month:
    uint priorDays = 0;
    for (uint m=1; m<date.month && (m<6 || date.month!=13); m++) {
      priorDays+= monthLength(m, yearLength(date.year));
    }
    if (leap(date.year) && date.month >= 6 && date.month != 13) {
      priorDays += 30;
    }
    // Add to that the absolute date of the last day of the preceding 
    // year and the day of the month:
    return roshHashanah(date.year) - 1 + priorDays + date.day;
  }
\end{verbatim}
\end{small}
The inverse conversion {\tt a2h} is defined by a similar sequence of computations.

The conversion between absolute and Gregorian dates is simpler, since every common year has 365 days and every leap year 366,
independent of any other consideration.  Thus, the absolute date corresponding to a given Gregorian date is easily computed once 
the Gregorian date of absolute day 1 is established as September 6, -3760.  This is derived from the established correspondence
between modern Hebrew and Gregorian dates.  (Note that in the absence of a Gregorian year 0, the year -3760 is conventionally
identified as 3761 B.C.E.)

Conversion between the Hebrew and Gregorian calendars is then performed by way of an intermediate computation of an absolute
date.  Successful conversion between the author's known Hebrew and Gregorian birthdays is seen as compelling evidence of program
correctness.

\section{Properties of the Calendar}\label{program}

The benefit of translation to ACL2 is the opportunity for formal verification of interesting program properties.  The proofs
presented in this section are formalized in the ACL2 script {\tt proof.lisp}.  Among these results is the critical property that
the length of every year, as determined by the prescribed postponements of Rosh Hashanah from the day of the molad, is one of
the admissible values that are accommodated by the allowed variations in the lengths of the months.  This fact is not even
mentioned by Maimonides in his seminal work of 1187, {\it Sanctification of the New Month} \cite{rambam}, which includes a
definitive specification of the dechiyot.  Various authors \cite{dershowitz, landau} have explained the $3^{\rm rd}$ and $4^{\rm th}$
dechiyot by describing situations in which they preclude an inadmissible year length, but these explanations fall short of a
comprehensive proof.

We shall present two separate proofs of the following result:

\begin{small}
\begin{verbatim}
  (defthm legal-year-lengths
    (implies (posp year)
             (member (yearlength year)
                     (if1 (leap year)
                          '(383 384 385)
                        '(353 354 355)))))
\end{verbatim}
\end{small}
The strategy of the first proof is to minimize the requirement of mathematical reasoning through computation:\medskip

{\sc Proof \#1}: We shall identify an integer $m$ such that the length of any year $y$ is the same as that of $y+m$.  The proof is
then completed by directly computing the length of each $y \in [1,m]$.

Note that the duration of a Metonic cycle is
\[
235\cdot {\cal L} = 235 \cdot (((29 \cdot 24) + 12) \cdot 1080 + 793) = 179876755 \mbox{ parts}
\]
and a week consists of $7 \cdot 24 \cdot 1080 = 181440$ parts.  We seek the smallest number $n$ of cycles that comprise an integral
number of weeks, i.e., the minimal $n$ such that $179876755n$ is divisible by 181440.  Consideration of the prime factorizations
$181440 = 2^6\cdot 3^4 \cdot 5 \cdot 7$ and $179876755 = 5 \cdot 47 \cdot 131 \cdot 5843$ leads to the solution
\[
n = 2^6\cdot 3^4 \cdot 7 = 36288.
\]
Let $m = 19n = 689472$ and $\tilde{y} = y + m$.  We shall show that the years $y$ and $\tilde{y}$ have the same length.

Let $y' = y + 1$ and $\tilde{y}' = \tilde{y} + 1$.  Let ${\cal M} = \{d,h,p\}$, $\widetilde{\cal M} = \{\tilde{d}, \tilde{h}, \tilde{p}\}$,
${\cal M}' = \{d',h',p'\}$, and $\widetilde{\cal M}' = \{\tilde{d}', \tilde{h}', \tilde{p}'\}$ be the delayed moladot of $y$, $\tilde{y}$,
$y'$, and $\tilde{y}'$, respectively.  By inspection of the function {\tt roshHa\-shanah}, the number of days of delay from $d$ to Rosh Hashanah
of year $y$ is determined by three parameters: (1) the position of $y$ in the Metonic cycle, $y \bmod 19$; (2) the day of the week of ${\cal M}$,
$d \bmod 7$; and (3) the time of day of ${\cal M}$, $h$:$p$.  The length of $y$, therefore, is determined by $y \bmod 19$, $y' \bmod 19$,
$d \bmod 7$, $d' \bmod 7$, $h$:$p$, $h'$:$p'$, and $d'-d$.

We shall show that the following conditions hold:

\begin{itemize}
\item [(1)] $y \bmod 19 = \tilde{y} \bmod 19$;
\item [(2)] $d \bmod 7 = \tilde{d} \bmod 7$;
\item [(3)] $h$:$p = \tilde{h}$:$\tilde{p}$;
\item [(4)] $\tilde{d}-d = 251827457$, a constant independent of $y$.
\end{itemize}
Since $y$ is arbitrary, it will then follow that the same relations hold between $y'$ and $\tilde{y}'$: $y' \bmod 19 = \tilde{y}' \bmod 19$,
$d' \bmod 7 = \tilde{d}' \bmod 7$, $h'$:$p' = \tilde{h}'$:$\tilde{p}'$, and $\tilde{d}' - d' = 251827457 = \tilde{d} - d$.  The last equation
implies
\[
d'-d = d' - d + (\tilde{d}' - d') - (\tilde{d} - d) = \tilde{d}'-\tilde{d},
\]
and according to the observation of the preceding paragraph, we may conclude that $y$ and $\tilde{y}$ have the same length.

The first condition holds trivially:
\[
\tilde{y} \bmod 19 = (y + 19n) \bmod 19 = y \bmod 19.
\]
To establish (2), (3), and (4), note that the difference between $\widetilde{\cal M}$ and ${\cal M}$ is
\[
1080(24(\tilde{d}\mbox{$-$}d) + (\tilde{h}\mbox{$-$}h)) + (\tilde{p}\mbox{$-$}p) =
25920(\tilde{d}\mbox{$-$}d) + 1080(\tilde{h}\mbox{$-$}h) + (\tilde{p}\mbox{$-$}p) \mbox{ parts}.
\]
Since $\widetilde{\cal M}$ and ${\cal M}$ are separated by $n$ 19-year cycles, this difference may be equated with
\[
179876755n = 6527367685440 = 25920\cdot 251827457 \mbox{ parts}.
\]
Thus,
\[
25920(\tilde{d} - d - 251827457) + 1080(\tilde{h}\mbox{$-$}h) + (\tilde{p}\mbox{$-$}p) = 0.
\]
Since
\[
|1080(\tilde{h}-h) + (\tilde{p}-p)| < 1080\cdot 23 + 1080 = 25920,
\]
we have $1080(\tilde{h}-h) + (\tilde{p}-p) = 0$, which implies $\tilde{h} = h$ and $\tilde{p} = p$, i.e., (3) is satisfied.
Furthermore, $\tilde{d} - d = 251827457 = 35975351\cdot 7$, which implies (2) and (4) .

As we have observed, it follows that the length of $y$ is the same as that of $\tilde{y} = y + m$.
The proof is completed by checking the length of each $y$, $1 \leq y \leq m = 689472$.~$\Box$\medskip

The computation of the length of each $y$ in the interval $[1,m]$ may be done simply by executing the ACL2 translation of the function
{\tt yearLength}.  However, for each $y$, this entails $y-1$ calls of the function {\tt MOLAD-LOOP-0} to compute the molad of $y$,
and another $y$ calls to compute that of $y+1$, with a total of
\[
\sum_{y=1}^m(2y-1) = m^2 = 475371638784.
\]
On a well-equipped MacBook Pro, the computation runs in approximately 8 hours.  Our implementation of the proof, therefore, uses a more
efficient algorithm, based on the recurrence formula {\tt molad-next} of Section~\ref{moladot}.
This provides a linear-time recursive computation of the $m$ year lengths, resulting in a reduced run time of 4 seconds.

Thus, in spite of its stated strategy, each of several steps of Proof \#1 requires significant effort.  Our second proof is
shorter, more enlightening, and surveyable.  It involves a detailed case analysis, but as seen in the proof
script, the case structure need not be explicitly supplied to the prover, which discovers it without assistance.\medskip

{\sc Proof \#2}: Once again, let ${\cal M} = \{d, h, p\}$ and ${\cal M}' = \{d', h', p'\}$ be the delayed moladot of years $y$ and $y+1$,
respectively.  Our objective is to show that the length of $y$, as determined by $d'-d$ and the $2^{\rm nd}$, $3^{\rm rd}$, and $4^{\rm th}$
dechiyot, is one of the specified values.

Suppose first that $y$ is a common year.  Then
\[
{\cal M}' = {\cal M} + 12\cdot {\cal L} = {\cal M} + \{354, 8, 876\}.
\]
If $h$:$p$ $<$ 15:204, then $d' = d + 354$ and the length of $y$ is admissible unless the $3^{\rm rd}$ dechiyah applies to either
${\cal M}$ or ${\cal M}'$ and none applies to the other.  But the $3^{\rm rd}$ does not apply to ${\cal M}$, and if it applies to
${\cal M}'$, then $d'$ is a Tuesday and since $354 = 50\cdot 7 + 4$, $d$ is a Friday and therefore subject to the $2^{\rm nd}$.

We may assume, therefore, that $h$:$p$ $\geq$ 15:204, which implies $d' = d + 355 = d + 50\cdot 7 + 5$ and $h'$:$p'$ $<$ 8:876.
Thus, neither the $3^{\rm rd}$ nor the $4^{\rm th}$ applies to ${\cal M}'$.  It follows that the length of $y$ is admissible
unless the $2^{\rm nd}$ applies to ${\cal M}'$ and none applies to ${\cal M}$.  But if $d'$ is a Wednesday or Friday, then $d$ is
a Friday or Sunday and the $2^{\rm nd}$ applies; if $d'$ is a Sunday, the $d$ is a Tuesday and the $3^{\rm rd}$ applies.

Now suppose $y$ is a leap year.  Then
\[
{\cal M}' = {\cal M} + 13\cdot {\cal L} = {\cal M} + \{353, 21, 589\}.
\]
If $h$:$p \geq$ 2:491, then $d' = d + 384$ and once again, the length of $y$ is admissible unless the $3^{\rm rd}$ applies to
either ${\cal M}$ or ${\cal M}'$ and none applies to the other.  But since $y$ is a leap year, the $3^{\rm rd}$ does not apply to
${\cal M}$.  If it applies to ${\cal M}'$, then $d'$ is a Tuesday, and since $384 = 54\cdot 7 + 6$, $d$ is a Wednesday and the
$2^{\rm nd}$ applies to ${\cal M}$.

Thus, we may assume $h$:$p <$ 2:491, which implies $d' = d + 383 = d + 54\cdot 7 + 5$ and $h'$:$p' \geq$ 21:589.  Since 
neither the $3^{\rm rd}$ nor $4^{\rm th}$ applies to ${\cal M}$, the length of $y$ is admissible unless the $2^{\rm nd}$ applies 
to ${\cal M}$ and none applies to  ${\cal M}'$.  But if $d$ is a Friday or Sunday, then $d'$ is a Wednesday or Friday and the
$2^{\rm nd}$ applies to ${\cal M}'$, and if $d$ is a Wednesday, then $d'$ is a Monday and the $4^{\rm th}$ applies.~$\Box$\medskip

Another well known property of interest pertains to the {\it keviyah}, or {\it character} of a year, which comprises the length 
of the year and the day of the week on which it begins.  Clearly, these parameters together determine the day of the week
corresponding to every date within the year.  Although there are 4 possible starting days and 6 year lengths, only 14 of the
$4 \times 6 = 24$ possible combinations are realized.  This was noted by Maimonides without explanation.
\begin{quote}
If the New Year's Day of a year, whether ordinary [common] or embolismic [a leap year], falls on a Tuesday, that year will always have
regular months; if the New Year's day of a year, whether ordinary or embolismic, falls on a Saturday or Monday, it will never have
regular months; if, however, it falls on a Thursday, we have to distinguish between the ordinary and the embolismic year; if it is an
ordinary year, it can never have defective months, according to this system of calculation; if it is an embolismic year, it can never
have regular months, according to this same system~\cite[Chapter VIII]{rambam}.
\end{quote}
His observation may be formalized as follows:

\begin{small}
\begin{verbatim}
  (defthm keviyot
    (implies (posp y)
             (let ((dw (dayofweek (roshhashanah y)))
                   (len (yearlength y)))
               (or (and (= dw 3) ;y begins on Tuesday
                        (member len '(354 384)))
                   (and (member dw '(0 2)) ;Saturday or a Monday
                        (member len '(353 355 383 385)))
                   (and (= dw 5) ;Thursday
                        (member len '(354 355 383 385)))))))
\end{verbatim}
\end{small}

{\sc Proof}: Of the 10 disallowed combinations, 9 are readily eliminated by the observation that Rosh Hashanah of $y+1$ cannot occur 
on Sunday, Wednesday, or Friday.  For example, if year $y$ begins on Thursday, then its length cannot be either 353 or 384, since this
would imply that $y+1$ begins on Sunday or Wednesday.

The single remaining case is a year of length 385 that begins on a Tuesday.  In this case, since
385 is divisible by 7, the following Rosh Hashanah must also be a Tuesday.  Using the same notation as in the preceding proofs,
recall that if the day $d$ of the molad ${\cal M}$ of $y$ is the preceding Monday, then $d'$ is either $d + 383$ (Saturday)
or $d + 384$ (Sunday), and according to the dechiyot, Rosh Hashanah of $y+1$ is Monday at the latest.  Therefore, $d$ must be Tuesday
and $d'$ is either $d + 383$ (Sunday) or $d + 384$ (Monday), and in the latter case, the time of ${\cal M'}$ must be earlier than
21:589.  In either case, Rosh Hashanah is Monday.~$\Box$\medskip

Our final result is a proof of Landau's claim (mentioned in Section~\ref{rh}) pertaining to the possible motivation for the
$1^{\rm st}$ dechiyah: the molad of every month occurs before the end of the first day of the month.  The proof is simple,
but involves extensive computation.

\begin{small}
\begin{verbatim}
  (defun firstofmonth (month year) 
    (as 'day 1 (as 'month month (as 'year year ()))))

  (defthm landau-thm
    (implies (and (posp year)
                  (posp month)
                  (<= month (monthsinyear year)))
             (<= (ag 'day (monthlymolad month year))
                 (h2a (firstofmonth month year)))))
\end{verbatim}
\end{small}

{\sc Proof}:  Each admissible year length is handled separately.  For each month of a year of a given length, we compute the number
of days elapsed from Tishri 1 to the first of the month (based on the month lengths determined by the year length) and the time elapsed
between the molad of Tishri and that of the month (i.e., the appropriate multiple of ${\cal L}$).  In the cases 355, 384, and 385,
the claim follows from the fact that the molad of Tishri occurs before noon on Tishri 1.

The same computation is inconclusive in the cases 353, 354, and 383, but in those cases, we succeed by using instead the
corresponding constraint on the molad of the following year.  For example, for a year of length 354, since the molad of the
following year occurs before noon on Tishri 1 and the molad of the year of interest is earlier by an interval $12\cdot {\cal L} =
\{354, 8, 876\}$, it follows that the earlier molad occurs before 7:204 on Tishri 1.  This improved bound is sufficient to establish
the claim.~$\Box$\medskip

\nocite{*}
\bibstyle{eptcs}
\bibliographystyle{eptcs}
\bibliography{calendar}

\begin{thebibliography}{1}
\providecommand{\bibitemdeclare}[2]{}
\providecommand{\surnamestart}{}
\providecommand{\surnameend}{}
\providecommand{\urlprefix}{\hspace{3.5 in} \linebreak Available at: }
\providecommand{\urlprefixx}{\hspace{2 in} \linebreak Available at: }
\providecommand{\urlprefixxx}{\hspace{3 in} \linebreak Available at: }
\providecommand{\url}[1]{\texttt{#1}}
\providecommand{\href}[2]{\texttt{#2}}
\providecommand{\urlalt}[2]{\href{#1}{#2}}
\providecommand{\doi}[1]{doi:\urlalt{http://dx.doi.org/#1}{#1}}
\providecommand{\eprint}[1]{arXiv:\urlalt{https://arxiv.org/abs/#1}{#1}}
\providecommand{\bibinfo}[2]{#2}

\bibitemdeclare{misc}{bromberg}
\bibitem{bromberg}
\bibinfo{author}{Irv \surnamestart Bromberg\surnameend}:
  \emph{\bibinfo{title}{Moon and the Molad of the {H}ebrew Calendar}}.
\newblock
  \urlprefixx\url{http://individual.utoronto.ca/kalendis/hebrew/molad.htm}.

\bibitemdeclare{unpublished}{ac}
\bibitem{ac}
\bibinfo{author}{Mentor~Graphics \surnamestart Corp.\surnameend}:
  \emph{\bibinfo{title}{Algorithmic {C} Datatypes}}.
\newblock \urlprefixxx\url{https://www.mentor.com/hls-lp/downloads/ac-datatypes}.

\bibitemdeclare{book}{dershowitz}
\bibitem{dershowitz}
\bibinfo{author}{Nachum \surnamestart Dershowitz\surnameend} \&
  \bibinfo{author}{Edward~M. \surnamestart Reingold\surnameend}
  (\bibinfo{year}{1997}): \emph{\bibinfo{title}{Calendrical Calculations}}.
\newblock \bibinfo{publisher}{Cambridge University Press}.

\bibitemdeclare{misc}{landau}
\bibitem{landau}
\bibinfo{author}{Remy \surnamestart Landau\surnameend}:
  \emph{\bibinfo{title}{Hebrew~Calendar~Science~and~Myths}}.
\newblock \urlprefixxx\url{https://hebrewcalendar.tripod.com/\#11}.

\bibitemdeclare{book}{rambam}
\bibitem{rambam}
\bibinfo{author}{Moses \surnamestart Maimoniodes\surnameend}
  (\bibinfo{year}{1956}): \emph{\bibinfo{title}{Mishneh Torah, Book 3, Treatise
  8: Sanctification of the New Month}}.
\newblock \bibinfo{publisher}{Yale University Press}.
\newblock \bibinfo{note}{Translated from Hebrew by Solomon Gandz, with
  introduction by Julian Obermann}.

\bibitemdeclare{inproceedings}{workshop}
\bibitem{workshop}
\bibinfo{author}{David~M. \surnamestart Russinoff\surnameend}
  (\bibinfo{year}{2020}): \emph{\bibinfo{title}{Formal Verification of
  Arithmetic {RTL}: Translating {Verilog} to {C++} to {ACL2}}}.
\newblock In: {\sl \bibinfo{booktitle}{ACL2 2020: 24th International Workshop
  on the ACL2 Theorem Prover and its Applications}}, \bibinfo{address}{held
  online}, \doi{10.4204/EPTCS.152.12}.
\newblock
  \urlprefix\url{http://acl2-2020.info/papers/formal-verification-of-arithmetic-rtl.pdf}.

\bibitemdeclare{book}{el}
\bibitem{el}
\bibinfo{author}{David~M. \surnamestart Russinoff\surnameend}
  (\bibinfo{year}{2022}): \emph{\bibinfo{title}{Formal Verification of
  Floating-Point Hardware Design: A Mathematical Approach}},
  \bibinfo{edition}{2nd} edition.
\newblock \bibinfo{publisher}{Springer}, \doi{10.1007/978-3-030-87181-9}.

\bibitemdeclare{book}{stern}
\bibitem{stern}
\bibinfo{author}{Sacha \surnamestart Stern\surnameend} (\bibinfo{year}{2001}):
  \emph{\bibinfo{title}{Calendar and Community}}.
\newblock \bibinfo{publisher}{Oxford University Press},
  \doi{10.1093/0198270348.001.0001}.

\end{thebibliography}
\end{document}